\documentclass[twocolumn,aps,prl,superscriptaddress,showpacs, byrevtex]{revtex4}
\usepackage{amsmath}
\usepackage{graphicx}
\pdfoutput=1

\begin{document}

\title{Broadband electromagnetic response and ultrafast dynamics \\
of few-layer epitaxial graphene}
\author{H. Choi}
\affiliation{Materials Sciences Division, E. O. Lawrence Berkeley National Laboratory, Berkeley, CA 94720}
\author{F. Borondics}
\affiliation{Advanced Light Source, E. O. Lawrence Berkeley National Laboratory, Berkeley, CA 94720}
\author{D. A. Siegel}
\affiliation{Materials Sciences Division, E. O. Lawrence Berkeley National Laboratory, Berkeley, CA 94720}
\affiliation{Department of Physics, University of California, Berkeley, CA 94720}
\author{S. Y. Zhou}
\affiliation{Materials Sciences Division, E. O. Lawrence Berkeley National Laboratory, Berkeley, CA 94720}
\affiliation{Department of Physics, University of California, Berkeley, CA 94720}
\author{M. C. Martin}
\affiliation{Advanced Light Source, E. O. Lawrence Berkeley National Laboratory, Berkeley, CA 94720}
\author{A. Lanzara}
\affiliation{Materials Sciences Division, E. O. Lawrence Berkeley National Laboratory, Berkeley, CA 94720}
\affiliation{Department of Physics, University of California, Berkeley, CA 94720}
\author{R. A. Kaindl}
\affiliation{Materials Sciences Division, E. O. Lawrence Berkeley National Laboratory, Berkeley, CA 94720}

\begin{abstract}
We study the broadband optical conductivity and ultrafast carrier dynamics of epitaxial graphene in the few-layer limit. Equilibrium spectra of nominally buffer, monolayer, and multilayer graphene exhibit significant terahertz and near-infrared absorption, consistent with a model of intra- and interband transitions in a dense Dirac electron plasma. Non-equilibrium terahertz transmission changes after photoexcitation are shown to be dominated by excess hole carriers, with a 1.2-ps mono-exponential decay  that reflects the minority-carrier recombination time.
\end{abstract}

\maketitle

The discovery of graphene -- a carbon monolayer and building block of graphite, fullerenes, and nanotubes -- provides unique opportunities to explore the properties of two-dimensional Dirac fermions \cite{Geim07}. The electromagnetic properties and ultrafast carrier dynamics, in particular, are important for applications of this new material \cite{Ryzhii2007,Mikhailov:2007PRL,Rana:2007PRB}. In exfoliated graphene, the infrared response is characterized by a universal quantum conductivity $\sigma_Q = \pi e^2 / 2 h$, arising from interband transitions whose onset energy follows the carrier density \cite{Nair:2008Science,Li:2008NPhys,Wang:2008Science,Mak:2008PRL}. A promising route towards large-scale device production is epitaxial growth of graphene layers on SiC substrates \cite{deHeer:2007SSC}. The optical response of epitaxial graphene is, however, much less explored. Recently, first equilibrium and time-resolved infrared and terahertz (THz) measurements were reported on epitaxial graphene with a large number ($N=6$--37) of layers \cite{Dawlaty:2008APLb,Dawlaty:2008APL,George:2008NanoLett,Sun:2008PRL}. In this letter, we present a broadband infrared and ultrafast THz study of {\it few-layer} epitaxial graphene. Systematic thickness variation covers nominally buffer, monolayer, and multilayer graphene films. We utilize equilibrium infrared spectroscopy to characterize the broadband conductivity and transient THz measurements to monitor the photoexcited carrier dynamics. This yields momentum and population relaxation times, and provides insight into graphene's unusual electrodynamics.

The samples studied here were grown via thermal decomposition on the Si-terminated face of semi-insulating 6H-SiC(0001) wafers \cite{Rollings:2006JPCS}. The film thickness and morphology was characterized {\it in situ} via low-energy electron microscopy (LEEM), showing nanoribbon-like monolayer terraces with widths of $\simeq 60$ to 250 nm. Angle-resolved photoemission spectroscopy (ARPES) and scanning tunneling microscopy (STM) evidenced a single-crystalline character of such domains interspersed with some defects \cite{Zhou:2007NMat,Rutter:2007Science}.

\begin{figure}[b!]
\centering \includegraphics[width=8.5cm]{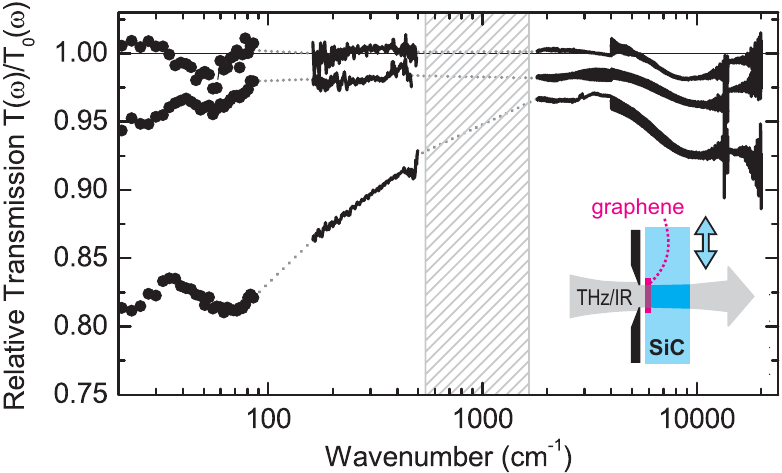}
\caption{(Color online) Relative transmission of nominally buffer, monolayer, and multilayer graphene (top to bottom), measured by FTIR (solid lines) and time-domain THz spectroscopy (dots). Dotted lines: guide to the eye. Undulations in the THz data below 100 cm$^{-1}$ arise from laser drift. Hatched: Reststrahlen region of SiC. Inset: sample modulation scheme.}
\end{figure}

Measurements of the equilibrium broadband infrared response from 20--2500 meV were carried out with Fourier-Transform Infrared (FTIR) spectroscopy, using a Bruker~IFS~66v spectrometer with four different source-detector combinations. This was complemented with  time-domain THz spectroscopy in the low-energy (3--10 meV) range~\cite{Kaindl:2009PRB}. The small absorption of the atomically-thin layers necessitates optimal suppression of systematic errors. We used rectangular SiC wafers ($5.5\times12$~mm), with graphene growth limited to a central, 4-mm diameter area via the heater and LEEM cap geometry. The sample was then spatially modulated below a 2-mm aperture in the spectrometer (inset, Fig.~1), alternating between graphene (transmission $T$) and the bare substrate ($T_0$). Systematic errors were thus reduced to about $\pm 1\%$ and $\pm 2\%$ for FTIR and THz spectra, respectively. Figure~1 shows the resulting relative transmission spectra $T(\omega)/T_{0}(\omega)$ at room temperature, for nominally buffer, monolayer, and multilayer graphene. A strong, systematic transmission decrease with increasing thickness is apparent, and all spectra show near-IR absorption above $\simeq 4000$~cm$^{-1}$. Moreover, monolayer and multilayer graphene feature far-IR absorption below $\simeq 500$~cm$^{-1}$, with considerable strength given the atomic-scale thickness.

More insight is obtained from the optical sheet conductivity $\sigma(\omega)$. It relates to the thin-film transmission via {\protect $T(\omega)/T_0(\omega) = |1+ Z_0 \sigma(\omega) / (n_{\rm S} +1)|^{-2}$}, where $Z_0$ is the vacuum impedance and $n_{\rm S}\simeq$ 2.5--3.1 the SiC refractive index \cite{Nuss:1991JAP,Dawlaty:2008APLb}. The influence of the imaginary part of conductivity is negligible for the given parameters. Thus, we can directly obtain the real part $\sigma_1(\omega)$ from the above expression, taking into account a frequency-dependent $n_{\rm S}$ \cite{Spitzer:1959PR}. Figure~2 shows the resulting sheet conductivity, normalized by $\sigma_Q$. The buffer layer response is insulating (Fig.~2a), as expected from a lack of Fermi-level electronic bands \cite{deHeer:2007SSC}, with some absorption around 1~eV. Figures~2(b) and 2(c) show the mono- and multilayer response after subtracting the buffer layer. It exhibits significant broadband conductivity, with a strong frequency dependence in multilayer graphene.

For further analysis, we calculate $\sigma_1(\omega)$ for an n-doped graphene layer, which at temperature $T$ reads
\begin{align}
\frac{\sigma_{1}(\omega)}{\sigma_Q} = & \frac{8k_{B}T}{\pi\hbar}
\text{ln}\left(e^{\frac{-E^e_{F}}{2k_B T}} + e^{\frac{E^e_{F}}{2k_B T}}\right) \frac{1}{\omega^2\tau+1/\tau}  \\ \nonumber
& + \frac{1}{2} \left[ \tanh\left( \frac{\hbar\omega+2E^e_{F}}{4k_B T} \right) + \tanh\left( \frac{\hbar \omega - 2E^e_{F}}{4k_B T} \right) \right],
\end{align}
\noindent where $\tau$ is the momentum scattering time and $E^e_F$ the electron Fermi energy~\cite{Stauber:2008PRB,Mikhailov:2007PRL}. The first term is the Drude-like intraband conductivity, while the second arises from interband transitions above~$\simeq2 E^e_F$. For high doping i.e. $E^e_F \gg 2k_B T$, the conductivity reduces to
\begin{equation}
\frac{\sigma_{1}(\omega)}{\sigma_Q} \simeq \frac{4 E^e_F}{\pi\hbar} \frac{1}{\omega^2\tau+1/\tau}
+ \left(1+\text{e}^\frac{2E_{F} - \hbar \omega}{2k_B T}\right)^{-1}.
\end{equation}
Here, $E^e_F \simeq \hbar v (\pi N)^{1/2}$, where $N$ is the electron density and $v \simeq 10^6$~m/s is the Dirac fermion velocity. Thus, unlike ordinary conductors, the Drude spectral weight is a {\it nonlinear} function of $N$.

The above model, scaled by the number of layers $n_L$, is shown as solid lines in Figs. 2(b) and 2(c). It provides for a faithful representation of the measured sheet conductivity, with $n_L = 1.5$, $E^e_F = 0.45$~eV, and $\tau = 2$~fs for the nominally monolayer sample, and $n_L = 4.5$, $E^e_F = 0.22$~eV, $\tau = 9.5$~fs for multilayer graphene. This model comparison shows a consistent scaling of the intraband spectral weight with the high-frequency, interband response set by $n_L \sigma_Q$. The Fermi energy reflects the large substrate-induced doping of few-layer graphene, corresponding to $N = 0.4$--$1.5\times10^{13}$~cm$^{-2}$, which compares well to ARPES and STM studies~\cite{Zhou:2007NMat,Rutter:2007Science}. We also calculated the conductivity with a model that includes the gap $2\Delta$ seen in ARPES \cite{Dawlaty:2008APLb,Zhou:2007NMat}. It is shown by the dashed lines in Figs.~2(b) and 2(c), with $2\Delta$=250 meV and 50 meV respectively. Clearly, at these doping levels the effect of the gap on $\sigma_1(\omega)$ is negligible. Figures~2(d)-(f) show LEEM images of all three samples, which underscore their nanoscale inhomogeneity \cite{Zhou:2007NMat}. Analysis of these images yields an average thickness of, respectively, 0.25, 1.1, and 3.3 ML. This agrees reasonably with the fits to the infrared spectra, when considering the limited LEEM field of view. The short scattering times from the model correspond to a mean-free path {$v_f\tau =$}~2--10~nm, well below the $\simeq$~100 nm graphene terrace size but consistent with scattering from impurities and inherent nanoscale ripples in graphene \cite{Rutter:2007Science}.

\begin{figure}[t!]
\centering
\includegraphics[width=8.5cm]{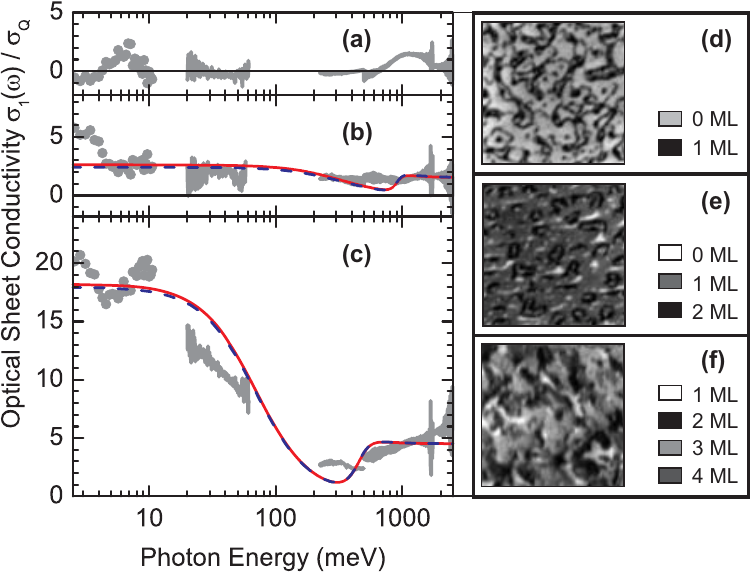}
\caption{(Color online) Real part of the optical sheet conductivity (gray), normalized by $\sigma_Q$: (a) buffer layer, (b-c) monolayer and multilayer graphene, with the buffer conductivity subtracted. Solid and dashed lines: model as explained in the text. (d)-(f) $2.5 \times 2.5~\mu$m$^2$ LEEM images of nominally buffer, monolayer, and multilayer graphene.}
\label{FIT}
\end{figure}

We now turn to the non-equilibrium THz dynamics measured via optical-pump THz-probe spectroscopy. The graphene layers are excited at room temperature with 1.53~eV femtosecond pulses from a 250-kHz Ti:sapphire regenerative amplifier and probed with ps THz pulses detected via electro-optic sampling \cite{Kaindl:2009PRB}. Figure~3(a) shows a typical electric field trace $E_0(t)$ transmitted through the unexcited sample, and the pump-induced change $\Delta E(t)$ measured at a fixed pump-probe delay $\Delta t$. The sign and negligible phase shift of $\Delta E(t)$ indicate a transmission decrease, i.e. added THz conductivity. Also, $\Delta E(t)$ decays in amplitude with increasing $\Delta t$ but retains its shape (not shown), allowing us to determine the overall field change at a fixed timepoint (arrow, Fig.~3a). The relative transmission change is $\Delta T / T_0 = 2\Delta E / E_0+ (\Delta E / E_0)^2$, which is plotted in Fig.~3(b). The signals peak within the time resolution after excitation, followed by a mono-exponential decay within the measurable range.

\begin{figure}[t!]
\centering \includegraphics[width=8.3cm]{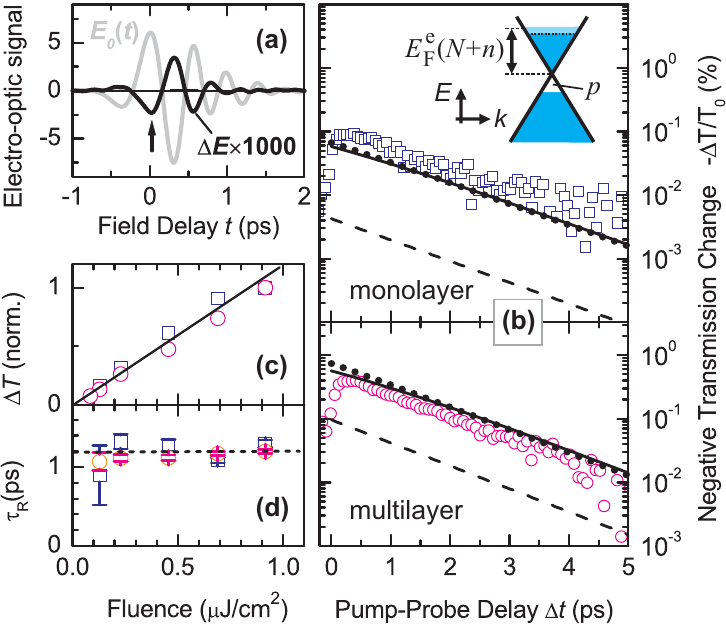}
\caption{(Color online) Non-equilibrium THz response. (a) Reference field $E_0$ (gray) and pump-induced change $\Delta E$ (black) at $\Delta t=0.4$~ps. (b) Transmission changes (symbols) at 0.9$~\mu$J/cm$^{2}$ pump fluence. Lines: model of Eq. 3, for electron (dashed) and hole contributions (dotted) at $T_c = 300$~K, and the sum (solid lines) with time-dependent $T_c$. Inset: non-equilibrium state. (c),(d) pump fluence dependence of amplitude $\Delta T$ and recombination time $\tau_R$ for mono- (squares) and multilayer (circles).}
\end{figure}

The maximum incident fluence of $0.9~\mu$J/cm$^2$ corresponds to photoexcited electron and hole densities $n_0 = p_0 = 4.6\times 10^{10}$cm$^{-2}$layer$^{-1},$ given the graphene interband absorption $2Z_0\sigma_Q/(n_S+1) \simeq 1.3\%$. After excitation, the excess carriers thermalize with the existing plasma on a fs timescale, forming a hot Fermi distribution with a temperature increased by $\Delta T$ above the lattice temperature $T_L$. We can estimate $\Delta T$ from energy conservation, $U_e(T_L,N) + \Delta Q = U_e(T_L + \Delta T, N + n) + U_h(T_L + \Delta T,p)$, where $U_{e,h}$ is the electron or hole gas internal energy and $\Delta Q$ is the absorbed pulse energy. For Dirac fermions $U_{e,h} = 4k_B^3T^3/(\pi v^2\hbar^2){\rm F_2}(E_F^{e,h}/k_BT)$, where F$_2$ is the second-order Fermi integral. For the highest fluence, this yields $\Delta T = 86$~K for the monolayer and 201~K for the multilayer sample. These are upper limits, as part of the energy is shed by phonon emission. The dense electron gas thus represents a heat bath, and the THz response can reflect changes of both majority and minority carriers (inset, Fig. 3b). This scenario is different from THz studies of thick multilayer graphene, dominated by photoexcited carriers in undoped layers \cite{George:2008NanoLett}.

For quantitative comparison, we calculate the non-equilibrium conductivity change
\begin{align}
\frac{\Delta\sigma_1(\omega)}{\sigma_Q} \simeq \left[ \frac{2 v}{\sqrt{\pi N}} \, n + \frac{4 k_B T_c}{\pi\hbar} \ln(1+e^{\frac{E_F^h(p)}{k_B T}}) \right]\frac{1}{\omega^2\tau+1/\tau},
\end{align}
for weak excitation $n \ll N$ and transient carrier temperature $T_c \ll E^e_F/2k_B$. This model is shown as lines in Fig.~3(b), assuming exponential decay $n=p=p_0 \exp(-t/\tau_R)$ with $\tau_R =$1.3~ps and 1.2~ps for mono- and multilayer graphene. Equation~3 includes two parts: $(i)$ the conductivity of photoexcited {\it electrons}, which follows from Eq.~2 with a non-equilibrium electron Fermi energy $E^e_F + \Delta E^e_F$. For weak excitation, $\Delta E^e_F \simeq (n/2N) E^e_F$ which renders the electron contribution {\it linear} in $n$. However, the electron response -- shown as dashed lines in Fig. 3(b) -- is about an order of magnitude too small, and thus fails to explain the observed signals. $(ii)$ The conductivity of photoexcited {\it holes} forms the second term in Eq.~3, which we derive from the full intraband expression using a hole distribution confined to the valence band \cite{Dawlaty:2008APLb,Gusynin:2006PRL}. It exhibits a generally {\it non-linear} dependence on the hole Fermi energy $E_F^h(p)$ and, likewise, on the hole density
\begin{align}
p(E_F^h,T_c) = \frac{-2k_B^2T_c^2}{\pi v^2 \hbar^2} {\rm Li_2}(-e^{E_F^h/k_BT_c}),
\end{align}
\noindent where Li$_2$ is the dilogarithm. The non-equilibrium hole response [dotted  lines, Fig.~3(b)] by far supersedes the electron contribution and yields a close description of the THz transmission change. This stark difference between electron and hole conductivity, evident in our highly-doped layers, reflects graphene's unusual sensitivity of the Drude spectral weight on the carrier distribution \cite{George:2008NanoLett}.

Three more aspects should be noted. First, the above evaluation assumed $T_c = T_L$. In contrast, the solid lines in Fig.~3(b) show the sum of electron and hole signals with a time-dependent carrier temperature $T_c = T_L + \Delta T \exp(-t/\tau_c)$, with a cooling time $\tau_c = 1.4$~ps as per Ref.~\onlinecite{Sun:2008PRL}. Clearly, the influence of cooling on the signals is minor. Second, stimulated THz emission due to interband transitions at the Dirac point has been predicted for photoexcited gapless graphene \cite{Ryzhii2007}. The lack of this effect in our signals reflects the presence of the electronic gap. Finally, as shown by the intensity dependence in Fig.~3(c), the response is nearly linear in $p$. This is confirmed by the model, see e.g. Fig.~3(b) where the hole response follows the electron curve. The linear hole response arises at sufficiently small densities where $E_F^h(p) < -kT_c$, corresponding to $p \lesssim 4 \cdot 10^{10}$cm$^{-2}$ at 300 K. In this limit, $p \simeq 2 k_B^2 T_c^2 \exp(E_F^h/k_BT_c)/(\pi v^2 \hbar^2)$ which renders the hole contribution in Eq.~3 proportional to $(2 v^2 \hbar/k_BT_c)\times p$.

The transmission changes in Fig.~3(b) thus represent the population dynamics of excess holes. Since few-layer epitaxial graphene is highly $n$-doped, electron-hole recombination is dominated by the interaction of the pump-induced minority carriers (holes) with the paramount electron plasma. This explains both the mono-exponential kinetics of the transient THz signals and the largely excitation-independent effective recombination time $\tau_R \simeq1.2$~ps in Fig. 3(d). This value of $\tau_R$ is consistent (within a factor of $\simeq3$) with calculations of Auger and phonon-mediated recombination \cite{Rana:2007PRB}.

In summary, we studied the broadband equilibrium conductivity of few-layer epitaxial graphene -- consistent with intra- and interband transitions of a dense Dirac electron plasma -- and measured the ultrafast minority-carrier recombination time via non-equilibrium THz transmission changes. Despite a balance of electron and hole excitations, the non-equilibrium THz response in these highly n-doped layers is shown to be dominated by holes, a confirmation of graphene's unusual electrodynamics.

\begin{acknowledgments}
This work was supported by the DOE Office of Basic Energy Sciences, Contract DE-AC02-05CH11231. F.~B. acknowledges a scholarship of the Rosztoczy Foundation.
\end{acknowledgments}









\begin{references}
\bibitem{Geim07} A. K. Geim and K. S. Novoselov, Nature Mat. {\bf 6}, 183 (2007).
\bibitem{Ryzhii2007} V. Ryzhii, M. Ryzhii, and T. Otsuji, J. Appl. Phys. {\bf101}, 083114 (2007).
\bibitem{Mikhailov:2007PRL} S. A. Mikhailov and K. Ziegler, Phys. Rev. Lett. {\bf99}, 016803 (2007); S. A. Mikhailov, Microelectron. J. {\bf 40}, 712 (2009).    
\bibitem{Rana:2007PRB} F. Rana, Phys. Rev. B. {\bf 76}, 155431 (2007); F. Rana, P. A. George, J. H. Strait, J. Dawlaty, S. Shivaraman, Mvs Chandrashekhar, and M. G. Spencer, Phys. Rev. B {\bf 79}, 115447 (2009).
\bibitem{Nair:2008Science} R. R. Nair, P. Blake, A. N. Grigorenko, K. S. Novoselov, T. J. Booth, T. Stauber, N. M. R. Peres, and A. K. Geim, Science {\bf 320}, 1308 (2008).
\bibitem{Li:2008NPhys} Z. Q. Li, E. A. Henriksen, Z. Jiang, Z. Hao, M. C. Martin, P. Kim, H. L. Stormer, and D. N. Basov, Nature Phys. {\bf 4}, 532 (2008).
\bibitem{Wang:2008Science} F. Wang, Y. B. Zhang, C. S. Tian, C. Girit, A. Zettl, M. Crommie, and Y. R. Shen, Science {\bf 320}, 206 (2008).
\bibitem{Mak:2008PRL} K. F. Mak, M. Y. Sfeir, Y. Wu, C. H. Lui, J. A. Misewich, and T. F. Heinz, Phys. Rev. Lett. {\bf 101}, 196405 (2008).
\bibitem{deHeer:2007SSC} W. A. de Heer, C. Berger, X. S. Wu, P. N. First, E. H. Conrad, X. B. Li, T. B. Li, M. Sprinkle, J. Hass, M. L. Sadowski, M. Potemski, and G. Martinez, Solid State Commun. {\bf 143}, 92 (2007).
\bibitem{Dawlaty:2008APL} J. M. Dawlaty, S. Shivaraman, J. Strait, P. George, M. Chandrashekhar, F. Rana, and M. G. Spencer, Appl. Phys. Lett. {\bf 92}, 042116 (2008).
\bibitem{Dawlaty:2008APLb} J. M. Dawlaty, S. Shivaraman, J. Strait, P. George, M. Chandrashekhar, F. Rana, M. G. Spencer, D. Veksler, and Y. Chen, Appl. Phys. Lett. {\bf 93}, 131905 (2008).
\bibitem{Sun:2008PRL} D. Sun, Z. Wu, C. Divin, X. Li, C. Berger, W. A. de Heer, P. First, and T. B. Norris, Phys. Rev. Lett. {\bf 101}, 157402 (2008).
\bibitem{George:2008NanoLett} P. A. George, J. Strait, J. Dawlaty, S. Shivaraman, M. Chandrashekhar, F. Rana, M. G. Spencer, Nano Lett. {\bf 8}, 4248 (2008).
\bibitem{Rollings:2006JPCS} E. Rollings, G.-H. Gweon, S. Y. Zhou, B. S. Mun, J. L. McChesney, B. S. Hussain, A. V. Fedorov, P. N. First, W. A. de Heer, and A. Lanzara, J. Phys. Chem. Solids {\bf 67}, 2172 (2006).
\bibitem{Zhou:2007NMat} S. Y. Zhou, G. H. Gweon, A. V. Fedorov, P. N. First, W. A. de Heer, D. H. Lee, F. Guinea, A. H. C. Neto, and A. Lanzara, Nature Mat. {\bf 6}, 770 (2007); D. A. Siegel, S. Y. Zhou, F. El Gabaly, A. V. Fedorov, A. K. Schmid, and A. Lanzara, Appl. Phys. Lett. {\bf 93}, 243119 (2008).
\bibitem{Rutter:2007Science} G. M. Rutter, J. N. Crain, N. P. Guisinger, T. Li, P. N. First, and J. A. Stroscio, Science {\bf 317}, 219 (2007).
\bibitem{Kaindl:2009PRB} R. A. Kaindl, D. H{\"{a}}gele, M. A. Carnahan, and D. S. Chemla, Phys. Rev. B {\bf 79}, 045320 (2009).
\bibitem{Nuss:1991JAP} M. C. Nuss, K. W. Goossen, J. P. Gordon, P. M. Mankiewich, M. L. O'Malley, and M. Bhushan, J. Appl. Phys. {\bf 70}, 2238 (1991).
\bibitem{Spitzer:1959PR} W. G. Spitzer, D. Kleinman, and D. Walsh, Phys. Rev. {\bf 113}, 127 (1959).
\bibitem{Stauber:2008PRB} T. Stauber, N. M. R. Peres, and A. K. Geim, Phys. Rev. B {\bf 78}, 085432 (2008).
\bibitem{Gusynin:2006PRL} V. P. Gusynin, S. G. Sharapov, and J. P. Carbotte, Phys. Rev. Lett. {\bf96}, 256802 (2006).
\end{references}
\end{document}